\documentclass[10pt]{article} 
\input epsf.tex
\def\ffrac#1#2{\textstyle{#1\over#2}\displaystyle}
\newcommand{\qed}{\hbox{\rule[-2pt]{3pt}{6pt}}}
\begin{document}
\begin{center}
{\LARGE \bf Families of Vicious Walkers}\\
 \vskip0.1in
{\large John Cardy}\\
Theoretical Physics, 1 Keble Road, Oxford OX1 3NP, United Kingdom\\
and All Souls College, Oxford\\
\vspace{2mm}
{\large Makoto Katori}\\
Department of Physics, Chuo University, Kasuga, Bunkyo-ku,
Tokyo 112-8551, Japan\\
\end{center}
\begin{abstract}
We consider a generalisation of the vicious walker problem in which
$N$ random walkers in ${\bf R}^d$
are grouped into $p$ families. Using field-theoretic renormalisation
group methods we calculate the asymptotic behaviour
of the probability that no pairs of walkers from different families 
have met up to time $t$. For $d>2$, this is constant, but for $d<2$ it
decays as a power $t^{-\alpha}$, which we compute to 
${\cal O}(\varepsilon^2)$ in an expansion in 
$\varepsilon=2-d$. The second order term depends on the ratios of the
diffusivities of the different families. In two dimensions, we find
a logarithmic decay $(\ln t)^{-\bar\alpha}$, and compute
$\bar\alpha$ exactly.
\end{abstract}
\section{Introduction}
\setcounter{equation}{0}
\label{sec1}
Consider the following problem: $N$ random walkers set off from the
vicinity of the origin, in $d$-dimensional euclidean space, at time
$t=0$. They are divided into $p$ different families: the number of
walkers in the $j$th family is $n_j$, so that $N=\sum_{j=1}^pn_j$.
Within a particular family, walkers are indifferent to each other:
their paths may cross. However, each family behaves viciously towards
all the others: if two walkers from different families meet,
both are annihilated. We may ask many different questions about this 
problem, but a fundamental quantity is
the probability $P(\{n_j\};t)$ that all the
walkers have still survived up to time $t$. Equivalently, we may
consider the ensemble of $N$ independent random walks: $P(\{n_j\};t)$
is the fraction of these in which no walkers of different families have
intersected up to time $t$. 

For a discrete time process on a lattice, 
if ${\bf r}^{\nu_j}_{j}(t)$ is the 
position at time $t$ of the $\nu_j$th walker of the $j$th family,
then $P(\{n_j\};t)$ is the expected value of the indicator function
\begin{equation}
\label{indic}
\prod_{t'=0}^t\prod_{1\leq j<k \leq p}\prod_{\nu_j=1}^{n_j}
\prod_{\nu_k=1}^{n_k}\left(1-\delta({\bf r}^{\nu_j}_{j}(t'),
{\bf r}^{\nu_k}_{k}(t'))\right)
\end{equation}

This problem is of interest for several reasons. It generalises
a number of cases:
\begin{enumerate}
\item $n_j=1$ ($1\leq j\leq p$) corresponds to \em vicious
walkers\em, a term first introduced by Fisher \cite{FisherVW}. It has
been studied using $\varepsilon$-expansion methods \cite{MB93a,MB93b}
similar to those of the present paper.
The survival probability is known exactly for $d=1$
in the case when all walkers have the same diffusion constants:
it decays as a power $t^{-p(p-1)/4}$ \cite{FisherVW,KGV00,KT01}. 
These methods rely essentially on a
fermionic description of the problem \cite{FisherVW,deGennes}.
Krattenthaler et al.\cite{KGV00} introduced the method
of the Schur functions and Katori and Tanemura \cite{KT01} developed this
and discussed the relation with the random matrix theory.\
These methods do not appear to extend
to the case when the diffusion constants are different.
Results in this case have been reported for $p=2$ \cite{FisherVW}.
\item The case $p=2$, with $n_1=n$ and $n_2=1$, has been studied by
Krapivsky and Redner\cite{KR96,RK99,R01} as a model of $n$ predators
(`lions') hunting a prey (`lamb'). They were able to obtain exact
results for the asymptotic behaviour of the survival probability, 
again in $d=1$, for the cases $n=1,2$ and arbitrary diffusion constants.
For general $n$, the exponent is related to the smallest eigenvalue of a
Dirichlet problem in a certain $(n-1)$-dimensional compact region, and is
probably not expressible analytically, but for large $n$ these authors
were able to estimate its behaviour.
The `lion-lamb' problem  for $d=1$
is related to a version of the `ballot problem' in which it is required
to know the probability that one candidate in a ballot remains ahead of
the $n$ others at all stages in the poll. Exact results are known only 
for $n\leq2$ \cite{Niederhausen83}.
\item The `lion-lamb' problem has another interpretation, in terms of
\em multiscaling\em: if we first regard the trajectory $\ell$ of the lamb
as fixed, and if $p_\ell(t)$ is the probability that it has not been met
by a \em single \em lion, then 
\begin{equation}
P(n,1;t)=\langle p_\ell(t)^n \rangle_\ell
\end{equation}
where the average is over all the realisations of $\ell$. 
The fact that $P(n,1;t)$ decays with $t$ with an exponent which is not
simply linear in $n$ is symptomatic of multiscaling in this problem.
\item More generally, we can regard $P(n_1,n_2,\ldots,n_p;t)$
as being the average of the $n_1$th power of the survival probability of
a \em single \em walker of family 1, in the presence of
$(n_2,\ldots,n_p)$ walkers of the $(p-1)$ other families.
\item Our problem has a strong resemblance to that of the intersection
probabilities of Brownian paths. In this case, one studies the ensemble
of $N$ random walks in $d$ dimensions each of
which begin a distance ${\cal O}(a)$ from the origin and which
arrive on a hypersphere of radius $r=R\gg a$ before they cross $r=a$,  
\em irrespective \em of how long this takes. Once again the walkers are
divided into families, and in this case one is interested in the
probability $\widetilde P(\{n_j\},R,a)$ that the \em paths \em of walkers
of different families do not intersect. Thus, instead of (\ref{indic}),
$\widetilde P$ is the expected value of
\begin{equation}
\prod_{1 \leq j<k \leq p}\prod_{\nu_j=1}^{n_j}
\prod_{\nu_k=1}^{n_k}\prod_{t'=0}^\infty\prod_{t''=0}^\infty
\left(1-\delta({\bf r}^{\nu_j}_{j}(t'),
{\bf r}^{\nu_k}_{k}(t''))\right)
\end{equation}
and it is supposed to decay as $(R/a)^{-\tilde\alpha}$ as
$R/a\to\infty$, where $\tilde\alpha$ depends nontrivially on the
$\{n_j\}$.
This problem is trivial in $d=1$, and turns out to have an upper
critical dimension $d=4$, below which an $\varepsilon$-expansion is
possible\cite{BPepsilon}. For $d=2$ an exact formula for
$\tilde\alpha(\{n_j\})$ has been derived\cite{Dup,LSW},
by exploiting the conformal invariance of the problem.
\end{enumerate}

Given these remarks, it seems important to investigate the general
case described in the opening paragraph. As far as we know, the 
fermionic methods
used to attack the vicious walker problem for $d=1$ do not extend to
this case. We have therefore employed a renormalisation group (RG)
method, which yields, for $d<2$, results for the exponent $\alpha(\{n_j\})$
of the power law decay of $P(\{n_j\};t)$ as a power series in
$\varepsilon\equiv2-d$. By using field-theoretic methods, the calculation
is streamlined, and, once the formalism is set up, involves relatively
little explicit calculation. We have carried this computation through 
${\cal O}(\varepsilon^2)$, and for arbitrary diffusion constants of each
family. It would be tedious, but not difficult, to carry it further,
as the actual Feynman integrals are elementary.
We also show that in two dimensions $P(\{n_j\};t)$ decays as a universal
power of $\ln t$.

The layout of this paper is as follows: in Sec.~\ref{sec2}, for
completeness, we collect all our results and show how they reduce in the
above-mentioned special cases. In Sec.~\ref{sec3} we set up the
field-theoretic formulation of the problem, then in the next section
carry out the RG analysis. Sec.~\ref{sec5} contains a summary and further
remarks. Several of the detailed calculations are relegated to
Appendices.

\setcounter{equation}{0}
\section{Results}
\label{sec2}
Let $p$ be the number of families, $n_j$ be the number of walkers in the
$j$th family, and $D_j$ be their diffusivity. 
Let $P(\{n_j\};t)$ be the survival probability
\begin{equation}
P(\{n_j\};t)=
E\left[\prod_{t'=0}^t\prod_{1\leq j<k \leq p}\prod_{\nu_j=1}^{n_j}
\prod_{\nu_k=1}^{n_k}\left(1-\delta({\bf r}^{\nu_j}_{j}(t'),
{\bf r}^{\nu_k}_{k}(t'))\right)\right]
\end{equation}
\subsection{$d>2$} In this case there is a finite probability that any
pair of walkers will never meet. As a result, 
$P(\{n_j\};t)$ approaches a non-universal constant value less than 1,
with leading power-law corrections of the form $t^{(2-d)/2}$.
\subsection{$d<2$}
\begin{equation}
P(\{n_j\};t)\sim {\rm const.}\,t^{-\alpha(\{n_j\})}
\quad{\rm as}\quad t\to\infty,
\end{equation}
where, with $\varepsilon=2-d$,
\begin{equation}
\alpha={\cal F}_{1} \varepsilon
+{\cal F}_{2} \ \varepsilon^2 
+{\cal O}(\varepsilon^3)
\label{eqn:result1}
\end{equation}
with 
\begin{eqnarray}
{\cal F}_{1} &=& \ffrac{1}{2} 
\sum_{1 \leq j_{1} < j_{2} \leq p} 
n_{j_{1}} n_{j_{2}} \nonumber\\
&=& \ffrac{1}{4} ( C_{1}^2-C_{2} ), \nonumber\\
{\cal F}_{2} &=& \ffrac{1}{2} 
\sum_{1 \leq j_{1} < j_{2} < j_{3} \leq p}
n_{j_{1}} n_{j_{2}} n_{j_{3}} 
\Big\{ \ln R(D_{j_{1}}, D_{j_{2}}, D_{j_{3}})
+ \ln R(D_{j_{2}}, D_{j_{3}}, D_{j_{1}})
+ \ln R(D_{j_{3}}, D_{j_{1}}, D_{j_{2}}) \Big\}
\nonumber\\
&+& \ffrac{1}{4} 
\sum_{1 \leq j_{1} < j_{2} \leq p}
n_{j_{1}} n_{j_{2}} 
\Big\{ (n_{j_{1}}-1) \ln R(D_{j_{1}}, D_{j_{2}}, D_{j_{1}})
+ (n_{j_{2}}-1)
\ln R(D_{j_{2}}, D_{j_{1}}, D_{j_{2}}) \Big\} \nonumber\\
&=& \ffrac{1}{2} \sum_{1 \leq j_{1} < j_{2} < j_{3} \leq p}
n_{j_{1}} n_{j_{2}} n_{j_{3}} 
\ln \left( \frac{(D_{j_{1}}D_{j_{2}}+D_{j_{1}}D_{j_{3}}
+D_{j_{2}}D_{j_{3}})^3}
{(D_{j_{1}}+D_{j_{2}})^2 (D_{j_{2}}+D_{j_{3}})^2
(D_{j_{3}}+D_{j_{1}})^2} \right) \nonumber\\
&+& \ffrac{1}{4} 
\sum_{1 \leq j_{1} < j_{2} \leq p}
n_{j_{1}} n_{j_{2}}
\left\{ n_{j_{1}} \ln \left( \frac{D_{j_{1}}(D_{j_{1}}+2D_{j_{2}})}
{(D_{j_{1}}+D_{j_{2}})^2} \right)
+ n_{j_{2}} \ln \left( \frac{(2D_{j_{1}}+D_{j_{2}})D_{j_{2}}}
{(D_{j_{1}}+D_{j_{2}})^2} \right)
\right\} \nonumber\\
&-& \ffrac{1}{4} 
\sum_{1 \leq j_{1} \leq j_{2} < p}
n_{j_{1}} n_{j_{2}} \ln \left(
\frac{D_{j_{1}}D_{j_{2}}(D_{j_{1}}+2D_{j_{2}})
(2 D_{j_{1}}+D_{j_{2}})}
{(D_{j_{1}}+D_{j_{2}})^4} \right), 
\label{eqn:F}
\end{eqnarray}
where
\begin{equation}
 C_{k}=\sum_{j=1}^{p} n_{j}^{k} \qquad
 k=1,2, \cdots,
\label{eqn:C}
\end{equation}
and
\begin{eqnarray}
R(D_{j}, D_{k}, D_{\ell})
&=& \frac{D_{j}D_{k}+D_{j}D_{\ell}+D_{k}D_{\ell}}
{(D_{j}+D_{k})(D_{k}+D_{\ell})}.
\label{eqn:R0}
\end{eqnarray}
From this may be deduced various special cases:
\subsubsection{Equal diffusion constants}
Assume that
$D_j=D$ for all $j=1,2,\cdots,p$.
Then
\begin{equation}
\alpha={\cal F}_{1} \varepsilon
+{\cal F}_{2} \ln \ffrac{3}{4} \ \varepsilon^2 
+{\cal O}(\varepsilon^3)
\label{eqn:result1b}
\end{equation}
with 
\begin{eqnarray}
{\cal F}_{1} &=& \ffrac{1}{2} 
\sum_{1 \leq j_{1} < j_{2} \leq p} n_{j_{1}} n_{j_{2}} \nonumber\\
&=& \ffrac{1}{4} \left( C_{1}^2-C_{2} \right), \nonumber\\
{\cal F}_{2} &=& 
\ffrac{3}{2} \sum_{1 \leq j_{1} < j_{2} < j_{3} \leq p}
n_{j_{1}} n_{j_{2}} n_{j_{3}} 
+\ffrac{1}{4} \sum_{1 \leq j_{1} < j_{2} \leq p}
n_{j_{1}}n_{j_{2}}(n_{j_{1}}+n_{j_{2}})
-\ffrac{1}{2} \sum_{1 \leq j_{1} < j_{2} \leq p}
n_{j_{1}} n_{j_{2}} \nonumber\\
&=& \ffrac{1}{4} \left(
C_{1}^3-C_{1}^2-2C_{1} C_{2} +C_{2}+C_{3} \right),
\label{eqn:F2}
\end{eqnarray}

Note that these are expressed in terms of symmetric polynomials in the 
$\{n_j\}$. This in fact holds to all orders in $\varepsilon$.
\subsubsection{Vicious walkers with unequal
diffusion constants}

When 
\begin{equation}
n_{j} =
\left\{
   \begin{array}{ll}
      1 & \quad \mbox{for} \quad 1 \leq j \leq p \\
      0 & \quad \mbox{otherwise}, \\
   \end{array}\right. \\
\label{eqn:vicious}
\end{equation}
$\alpha$ should be equal to the 
{\it survival exponent} $\psi_{S,p}$ of the 
vicious walkers.
In this case
$C_k=p$ for $k=1,2,\cdots$
and the result (\ref{eqn:result1}) gives 
\begin{eqnarray}
\psi_{S, p} &=& 
\left.\alpha \right|_{n_{j}=1 \ ( 1 \leq j \leq p), \
n_{k}=0 \ (k \geq p+1)} \nonumber\\
&=& \ffrac{1}{2} { p \choose 2} \varepsilon
+\ffrac{1}{2} 
\sum_{1 \leq j_{1} < j_{2} < j_{3} \leq p}
\ln \left( \frac{(D_{j_{1}}D_{j_{2}}+D_{j_{1}}D_{j_{3}}
+D_{j_{2}}D_{j_{3}})^3}
{(D_{j_{1}}+D_{j_{2}})^2 (D_{j_{2}}+D_{j_{3}})^2
(D_{j_{3}}+D_{j_{1}})^2} \right)
\ \varepsilon^2
+{\cal O}(\varepsilon^3).\nonumber\\
\label{eqn:result2}
\end{eqnarray}
\subsubsection{Vicious walkers
with equal diffusion constants}

The result (\ref{eqn:result1b}) gives 
\begin{eqnarray}
\psi_{S, p} &=& 
\left.\alpha \right|_{n_{j}=1 \ ( 1 \leq j \leq p), \
n_{k}=0 \ (k \geq p+1)} \nonumber\\
&=& \ffrac{1}{2} { p \choose 2} \varepsilon
+\ffrac{3}{2} {p \choose 3} \ln \ffrac{3}{4} 
\ \varepsilon^2
+{\cal O}(\varepsilon^3)
\nonumber\\
&=& \ffrac14p(p-1)  \varepsilon 
+\ffrac14p(p-1)(p-2) \ln \ffrac{3}{4} 
\ \varepsilon^2
+{\cal O}(\varepsilon^3).
\label{eqn:result2b}
\end{eqnarray}
This agrees with the result reported as Eqn.~(5.2),
with Eqn.~(3.13) in Mukherji and Bhattacharjee\cite{MB93a} 
(see also \cite{MB93b}).

It has been proved that \cite{FisherVW,KGV00,KT01}
\begin{equation}
\psi_{S, p}= \ffrac14p(p-1) \qquad
\mbox{for} \quad d=1 \quad (i.e. \quad
\varepsilon=1).
\label{eqn:VW}
\end{equation}
Note that although this exact result agrees with that from the
first-order $\varepsilon$-expansion (\ref{eqn:result2b}) on setting
$\varepsilon=1$, this is probably fortuitous, as, in the case of unequal
diffusivities the exact result depends on their ratio, while the
first-order term in (\ref{eqn:result2}) does not.
\subsubsection{`Lion-lamb' problem
with unequal diffusion constants}

The `$n$ lions and one lamb' problem studied by
Krapivsky and Redner \cite{KR96,RK99} is a special
case of the present model in which
\begin{equation}
n_{j} =
\left\{
   \begin{array}{ll}
      n & \quad \mbox{for} \quad j=1 \\
      1 & \quad \mbox{for} \quad j=2 \\
      0 & \quad \mbox{otherwise}. \\
   \end{array}\right. \\
\label{eqn:vicious2}
\end{equation}
In this case
$C_k=1+n^k$ for $k=1,2,\cdots$
and the result (\ref{eqn:result1}) gives 
\begin{eqnarray}
\beta_{n} &=& 
\left.\alpha \right|_{n_{1}=n, n_{2}=1, n_{j}=0 \ (j \geq 3)} 
\nonumber\\
&=& \ffrac12n  \varepsilon 
+ \ffrac14n(n-1)
\ln \left( \frac{1+2 \eta}{(1+\eta)^2}
\right)
\ \varepsilon^2
+{\cal O}(\varepsilon^3),
\label{eqn:result3}
\end{eqnarray}
where
$\eta=D_2/D_1$.
Redner and Krapivsky \cite{RK99} reported the exact solution
for $n=2$ in $d=1$ (i.e., $\varepsilon=1$),
\begin{equation}
\beta_{2}^{\rm exact}(\eta)
=\left[ 2 - \frac{2}{\pi} \cos^{-1} \frac{\eta}{1+\eta} \right]^{-1}.
\label{eqn:RK1}
\end{equation}
It was shown that $\beta^{\rm exact}(\eta)$ is monotonically
decreasing in $\eta$ and
\begin{equation}
\beta_{2}^{\rm exact}(0)=1, \quad
\beta_{2}^{\rm exact}(1)=\ffrac{3}{4}, \quad
\lim_{\eta \to \infty} \beta_{2}^{\rm exact}(\eta)=\ffrac{1}{2}.
\label{eqn:BK2}
\end{equation}
If we neglect ${\cal O}(\varepsilon^3)$ and set $n=2, \varepsilon=1$
in (\ref{eqn:result3}), we have
$$
\beta_{2}^{\rm approx.}(\eta)=1+\ffrac{1}{2}
\ln \left( \frac{1+2\eta}{(1+\eta)^2} \right),
$$
which is monotonically decreasing in $\eta$ and
$$
\beta_{2}^{\rm approx.}(0)=1, \quad
\beta_{2}^{\rm approx}(1)=1+ \ffrac{1}{2} \ln \ffrac{3}{4}
\simeq 0.856, \quad
\lim_{\eta \to \infty} \beta_{2}^{\rm approx.}(\eta)=-\infty.
$$

\subsection{Two dimensions}
In this case, there is a logarithmic decay with universal exponent:
\begin{equation}
\label{2dlog}
P(\{n_j\};t)\sim {\rm const.}\,(\ln t)^{-\bar\alpha}\left(1+
{\cal O}\left({1\over\ln t}\right)\right)
\end{equation}
where 
\begin{equation}
\bar\alpha=\sum_{1\leq j_{1} < j_{2} \leq p}
n_{j_{1}}n_{j_{2}}
\end{equation}
Note that this is independent of the $D_j$ (as long as no pair of them
both vanish): the dependence shows up only in the prefactor and the
non-leading terms.

\setcounter{equation}{0}

\section{Field-theoretic formulation}
\label{sec3}
In this section, we set up the general problem as a continuum field theory,
so that the powerful techniques of the field-theoretic RG may be
applied. The general method for formulating such stochastic particle systems
as field theories, as originally proposed by Doi\cite{Doi} and 
Peliti\cite{Peliti}, has been described at length elsewhere\cite{Car99}
and we shall only summarise how this is applied in the case of interest.

Initially, the problem is formulated on a lattice, for example
${\bf Z}^d$, the sites of which are labelled by a vector $\bf r$. 
The microstate of the system at a given time is specified by occupation
numbers $\{m_j({\bf r})\}$, which specify that there are $m_j({\bf r})$
walkers of family $j$ at the site $\bf r$. Note that we treat walkers
of the same family as identical particles: this makes no difference
in the problem of interest. To each microstate is associated a vector in
a Fock space $\cal F$, built by applying raising operators to the vacuum, or
empty state, $|0\rangle$:
\begin{equation}
|\{m_j({\bf r})\}\rangle=
\prod_{{\bf r}\in{\bf Z}^d}\prod_{j=1}^p{a_j^\dag({\bf r})}^{m_j({\bf
r})}|0\rangle\quad,
\end{equation}
where $[a_j({\bf r}),a_{k}^\dag({\bf r}')]=\delta_{jk}\delta_{{\bf r}
{\bf r}'}$, and $a_j({\bf r})|0\rangle=0$.
Let $p(\{m_j({\bf r})\};t)$ be the probability of finding the system in
this microstate at time $t$, and define the state $\in{\cal F}$
\begin{equation}
|\Psi(t)\rangle\equiv\sum_{\{m_j({\bf r})\}}p(\{m_j({\bf r})\};t)
|\{m_j({\bf r})\}\rangle\quad.
\end{equation}

Then the master equation, which is linear equation describing the
time-evolution of the probabilities $p(\{m_j({\bf r})\};t)$, 
is equivalent to the
Schr\"odinger-like equation
\begin{equation}
d|\Psi(t)\rangle/dt=-\hat H|\Psi(t)\rangle
\end{equation}
where $\hat H:{\cal F}\to{\cal F}$ may be expressed explicitly in terms of
the raising and lowering operators. For the case of independent random
walks in continuous time,
\begin{equation}
\hat H=\hat H_0=\sum_{j=1}^p(D_j/b^2)
\sum_{({\bf r},{\bf r}')}\big(a_j^\dag({\bf r})-
a_j^\dag({\bf r}')\big)\big(a_j({\bf r})-a_j({\bf r}')\big)
\end{equation}
where $b$ is the lattice spacing, and the sum is over nearest neighbour
pairs of sites $({\bf r},{\bf r}')$.

The probability of finding the walkers at sites ${\bf r}_j^{\nu_j}$
at time $t$
(where $1\leq\nu_j\leq n_j$ with $1\leq j\leq p$) is then given by
\begin{equation}
\label{prob}
\langle0|\prod_{j=1}^p\prod_{\nu_j=1}^{n_j}
a_j({\bf r}_j^{\nu_j})\,{\rm e}^{-t\hat H_0}\,|\Psi(0)\rangle
\end{equation}
Of course, when this is summed over all the ${\bf r}_j^{\nu_j}$, it
gives unity.

Before considering how to implement the non-intersection constraint,
let us first discuss the continuum limit and the path integral
representation. In this non-interacting case, the continuum limit may be
taken rigorously. The raising and lowering operators go over into 
(distribution-valued) field operators satisfying
$[\phi_j({\bf r}),\phi_{k}^\dag({\bf r}')]=\delta_{jk}\delta({\bf
r}-{\bf r}')$, and the generator of time evolution becomes
\begin{equation}
\hat H_0=\int\left[\sum_{j=1}^pD_j(\nabla\phi_j^\dag)(\nabla\phi_j)\right]d^d\!r
\end{equation}
Since the walkers are all supposed to begin in the vicinity of the
origin at $t=0$, that is, a finite number of lattice spacings away,
in the continuum limit $b\to0$
\begin{equation}
|\Psi(0)\rangle={\cal O}^\dag|0\rangle\equiv\prod_{j=1}^p
\left(\phi_j^\dag({\bf 0})\right)^{n_j}|0\rangle
\end{equation}

The path integral representation is derived\cite{Car99} by breaking the
time interval $(0,t)$ into slices of length $\Delta t$, so that the
time-evolution operator ${\rm e}^{-t\hat H_0}$ is the product of factors
${\rm e}^{-\Delta t\hat H_0}\approx 1-\Delta t\hat H_0$, and inserting a
complete set of coherent states at each time slice. This has the effect
of replacing the operators $\phi_j({\bf r})$ and $\phi_j^\dag({\bf r})$
by time-dependent $c$-number fields $\phi_j(t,{\bf r})$ and
$\phi_j^*({t,\bf r})$ respectively. After taking the limit
$\Delta t\to0$, the matrix element (\ref{prob})
becomes a functional integral
\begin{equation}
\label{pathintegral}
\int\prod_{j=1}^p{\cal D}\phi_j^*{\cal D}\phi_j
\prod_{j=1}^p\prod_{\nu_j=1}^{n_j}\phi(t,{\bf r}_j^{\nu_j}){\cal O}^*(0,
{\bf 0})\,{\rm e}^{-S_0}
\end{equation}
where 
\begin{equation}
S_0=\int\left[\sum_{j=1}^p\phi^*\partial_t\phi-
\sum_{j=1}^pD_j(\nabla\phi_j^*)(\nabla\phi_j)\right]dtd^d\!r
\end{equation}
and ${\cal O}^*=\prod_{j=1}^p{\phi^*_j}^{n_j}$.

\subsection{Interactions}

We now discuss how to incorporate the constraint that walkers of
different families should not meet. 
Rather than insert the indicator function (\ref{indic}) into the path
integral, it more convenient to consider a slightly more general problem
in which, before taking the limits $b\to0$ and $\Delta t\to0$, each
set of trajectories is weighted by a factor
\begin{equation}
\prod_{t'=0}^t\prod_{\bf r}
\exp\left(-\sum_{1\leq j_{1} < j_{2} \leq p}
\lambda_{j_{1}j_{2}}(b^d/\Delta t)
m_{j_{1}}(t',{\bf r})m_{j_{2}}(t',{\bf r})\right)
\end{equation}
where the $\lambda_{j_{1}j_{2}}(b^d/\Delta t)>0$ 
are a set of dimensionless
parameters (the factors of $b$ and $\Delta t$ are inserted to make the
continuum limit simpler.) The case of strict non-intersection
corresponds to the limit 
$\lambda_{j_{1}j_{2}}\to\infty$. However, we shall
show that, for $d\leq2$, the leading behaviour is independent of
the precise value of these parameters (as long as they are all strictly
positive) and, moreover, the RG fixed point, at which non-leading
corrections to the asymptotic behaviour disappear, corresponds to the
limit of infinite $\lambda_{j_{1}j_{2}}$. 

In the formal continuum limit, this corresponds to a modification of
the action in the path integral
\begin{equation}
\label{Sint}
S=S_0+\sum_{1\leq j_{1} < j_{2} \leq p}
\lambda_{j_{1}j_{2}}
\int\phi_{j_{1}}^*\phi_{j_{2}}^*\phi_{j_{1}}\phi_{j_{2}}dtd^d\!r
\end{equation}

\subsection{Feynman rules}
\begin{figure}
\centerline{
\epsfxsize=3in
\epsfbox{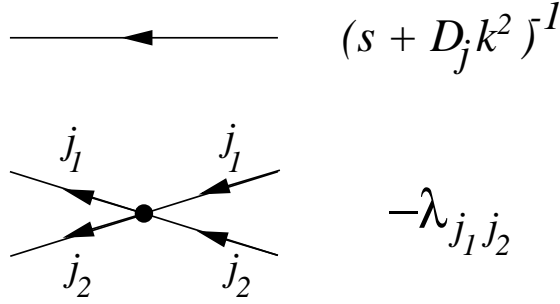}}
\caption{Feynman rules for the interacting theory (\ref{Sint}).
Time always flows towards the left.}
\label{figrules}
\end{figure}

The Feynman rules for this theory are very simple\cite{Car99}
and are illustrated in Fig.~\ref{figrules}.
We denote averages and correlations with respect to the bare action $S_0$ 
by the subscript ${}_0$: averages with respect to the full action $S$ are
denoted by $\langle\cdot\rangle$. 
\begin{itemize}
\item the Fourier-Laplace transform of the bare propagator 
\begin{eqnarray}
G^{(1,1)}_j(s,{\bf k})_0&=&\int_0^\infty dt\int d^d\!r e^{-st}
e^{i{\bf k}\cdot{\bf r}}\langle\phi_j(t,{\bf r})\phi_j^*(0,{\bf
0})\rangle_0\nonumber\\
&=&(s+D_jk^2)^{-1}
\end{eqnarray}
is represented by a line directed towards increasing time
(conventionally, right-to-left). In the $(t,{\bf k})$ representation
the bare propagator is simply ${\rm e}^{-D_jk^2t}$;
\item the interaction $(-\lambda_{j_1j_2})$ is represented by a vertex
with one incoming and one outgoing pair of
lines each of type $j_1$ and $j_2$;
\item
As usual, wave number $k$ and (imaginary) frequency $s$ are conserved
at the vertices, and internal loop integrations
$\int(ds/2\pi i)$ and $\int(d^d\!k/(2\pi)^d)$ are carried out.
\end{itemize}

\subsection{Renormalisation and operator product expansion}
The survival probability is now given by the correlation function
\begin{equation}
\overline G_{\cal O}(t)
=\int\prod_{j=1}^p\prod_{\nu_j=1}^{n_j}
\langle\phi_j(t,{\bf r}_j^{\nu_j}){\cal O}^*(0,{\bf 0})\rangle\quad,
\end{equation}
evaluated with the weight ${\rm e}^{-S}$.

However, this does not exist in the formal continuum limit,
 because the perturbative
Feynman diagram expansion of $\overline G$ contains ultraviolet
(short-distance or short-time) divergences. Physically
this is because two walkers, having interacted once, are, in the
continuum limit, likely to interact an infinite number of times
as $\Delta t\to0$. This divergence may be regulated, either by imposing
an explicit cut-off $|k|<\Lambda$ in the Feynman integrals, or, more
easily, by dimensional regulation. For $d\leq2$ this field theory 
is renormalisable: the singular dependence on the regulator may be
absorbed into a finite number of parameters. In the case of the theory
of interest, this procedure is particularly simple\cite{Car99}:
no renormalisation of the field $\phi(t,{\bf r})$ nor of the
diffusion constants is required, only a
simple renormalisation of the coupling constants $\lambda_{jj'}$, which
can be computed exactly to all orders.
The lack of field and diffusion constant renormalisation holds
mathematically because there are no loop corrections to the propagators.
Physically it is because an isolated walker does not interact, even with
itself, in the absence of any branching processes.
When the coupling constant renormalisation is done, all
correlation functions of products of $\phi$ and $\phi^*$, at 
\em distinct \em space-time points, have a finite limit as the regulator
is removed, when expressed in terms of the renormalised couplings.
The fact that the renormalised theory must be defined at some arbitrary
scale then leads to RG equations. 

However, this procedure is not sufficient to render finite correlation
functions involving so-called \em composite operators 
\em like ${\cal O}^*=\prod_{j=1}^p\phi_j^*(0,{\bf 0})^{n_j}$. 
Physically, this is because
if the walkers all begin at exactly the same point, they will all
annihilate each other immediately!
In order to obtain finite renormalised correlation functions, it is
first necessary to point-split the fields:
\begin{equation}
\prod_{j=1}^p\phi_j^*(0,{\bf 0})^{n_j}\longrightarrow
\prod_{j=1}^p\phi_j^*(0,{\bf r}_j)^{n_j}
\end{equation}
(Note that it is not necessary to split the starting points of walkers
of the same family, since they do not interact.) 
Now consider a correlation function of this product with an
arbitrary product $A$ of fields whose time arguments are 
all strictly positive:
\begin{equation}
\langle A \prod_{j=1}^p\phi_j^*(0,{\bf r}_j)^{n_j}\rangle
\end{equation}
In the cut-off theory, we could simply make a Taylor expansion
of this in powers of the ${\bf r}_j$. 
This would have the form 
\begin{equation}
\label{taylor}
\langle A{\cal O}^*(0,{\bf 0})\rangle 
+ \sum_nC_n(\{{\bf r}_j\})\langle A
{\cal O}^*_n(0,{\bf 0})\rangle
\end{equation}
where the summation is taken over all possible derivatives 
$\{{\cal O}^*_n\}$ of 
$\prod_j\phi_j^*(0,{\bf r}_j)^{n_j}$ with respect to the $\{{\bf
r}_j\}$.
On the basis of dimensional
analysis, the first term gives the leading behaviour for 
$\overline G_{\cal O}(t)$
as $t\to\infty$, at least at $d=2$. 
In the interacting theory, however, each term in (\ref{taylor}) has to
be renormalised separately. As a result
\begin{equation}
\label{OPE0}
\langle A \prod_{j=1}^p\phi_j^*(0,{\bf r}_j)^{n_j}\rangle=
Z_{\cal O}^{-1}\langle A{\cal O}^*_R\rangle+
\sum_n Z_{{\cal O}_n}^{-1}
C_{n,R}(\{{\bf r}_{j}\})\langle A{\cal O}^*_{n,R}\rangle
\end{equation}
where all the correlation functions are finite as the regulator is removed.
Since each of the $\{{\cal O}_{n,R}\}$ may acquire a nontrivial
anomalous dimension through this procedure, the renormalised functions
$C_{n,R}(\{{\bf r}_{j}\})$ have a non-trivial dependence on their
arguments.
The important feature of (\ref{OPE0}) is that the
renormalisation constants $Z_{\cal O}, Z_{{\cal O}_n}$ are
independent of $A$. This we may write the \em operator product 
expansion \em (OPE) 
\begin{equation}
\label{OPE}
\prod_{j=1}^p\phi_j^*(0,{\bf r}_j)^{n_j}
=Z_{\cal O}^{-1}{\cal O}^*_R(0,{\bf 0})
+\sum_n Z_{{\cal O}_n}^{-1}
C_{n,R}(\{{\bf r}_{j}\}){\cal O}^*_{n,R}(0,{\bf 0})
\end{equation}
where the OPE functions $C_{n, R}(\{{\bf r}_{j}\})$, \em etc. \em 
are in general nontrivial.

Each term in the OPE (\ref{OPE}), when substituted into $\overline G$,
will give rise to nontrivial power-law dependence on $t$ for $d\leq2$.
However, since in the noninteracting theory we know that only the first
term is important as $t\to\infty$, we shall assume that this remains
true for sufficiently small $\varepsilon$. Further discussion of this
point will be postponed to Sec.~\ref{sec5}. 

\subsection{$d>2$}
\begin{figure}
\centerline{
\epsfxsize=4in
\epsfbox{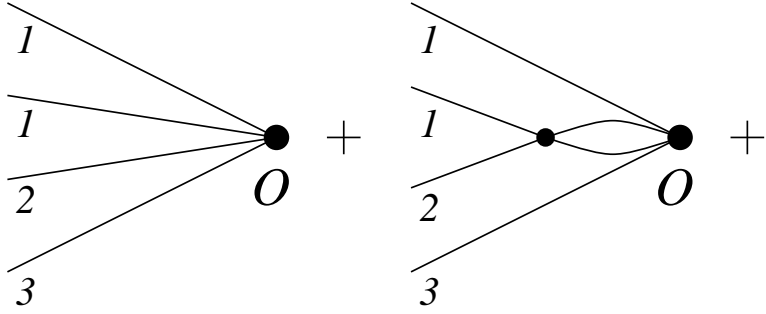}}
\caption{Some of the diagrams contributing to the survival probability
$\overline G_{\cal O}$. The case $n_1=2$, $n_2=1$, $n_3=1$ is shown as
an illustration. The one-loop diagram shows an interaction between a
walker of family 1 with one of family 2, and is proportional to
$\lambda_{12}$.}
\label{figGbar}
\end{figure}

In the absence of any interactions, the survival probability
$\overline G_{\cal O}(t)=1$, as can be seen by evaluating the first 
diagram in the expansion shown in Fig.~\ref{figGbar} with all 
the wave numbers ${\bf q}_j^{\nu_j}$ set to ${\bf 0}$. 
For $d>2$, the
higher order terms give contributions which correct this constant, as
well as terms which are subleading as
$t\to\infty$. For example, a typical one-loop diagram like that in
Fig.~\ref{figGbar} is proportional to
\begin{equation}
-\lambda_{j_{1}j_{2}}\int_0^tdt'\int d^d \! k\, 
{\rm e}^{-(D_{j_{1}}+D_{j_{2}}){\bf k}^2t'}
\propto
-\int{1-{\rm e}^{-(D_{j_{1}}+D_{j_{2}}){\bf k}^2t}
\over {\bf k}^2}d^d\!k
\end{equation}
This integral diverges at large $|{\bf k}|$ for $d\geq2$, a
consequence of taking the naive continuum limit. If we impose a cut-off
$|{\bf k}|<\Lambda={\cal O}(b)$, the leading term behaves as $\Lambda^{d-2}$,
with a non-universal coefficient, while the remainder is finite as
$\Lambda\to\infty$ and behaves as $t^{-(d-2)/2}$. 
The non-universal constant term
corresponds to a finite probability that the walkers
annihilate at short times, before escaping each other.
This behaviour persists at higher orders in the interactions.

For $d\leq2$, however, each successive term in the bare perturbation
grows as a larger and larger positive power of $t$, and it is necessary
to resum the expansion. The renormalisation group (RG) provides a
consistent framework within which to carry this out.
\setcounter{equation}{0}
\section{Renormalisation Group Analysis}
\label{sec4}
\subsection{Coupling constant renormalisation}
As usual, the renormalised couplings $\lambda_{Rjk}$
are defined as the values at the normalisation point
of the irreducible vertex functions
$\Gamma_{jk}^{(2,2)}\big((s_j',{\bf q}_j'),(s_k',{\bf q}_k')
;(s_j,{\bf q}_j),(s_k,{\bf q}_k)\big)$, which are the truncated
Laplace-Fourier transforms of 
$\langle\phi_j(t_j',{\bf r}_j')\phi_j(t_k',{\bf r}_k')\phi_j^*(t_j,{\bf
r}_j)\phi_j^*(t_k,{\bf r}_k)\rangle$.
(There is no field renormalisation in this theory.)
It is convenient to choose the normalisation point as 
\begin{equation}
s_j'=s_k'=s_j=s_k=\sigma\not=0;\qquad
{\bf q}_j'={\bf q}_k'={\bf q}_j={\bf q}_k=0
\end{equation}
This class of theories has the special property that the renormalised
coupling constants may be computed to all orders\cite{Car99}. The
calculation is summarised in Appendix~A. The result is
\begin{equation}
\label{CCR}
\lambda_{Rjk}={\lambda_{jk}\over
1+(b_d/\varepsilon)\lambda_{jk}((D_j+D_k)/2)^{-d/2}(2\sigma)^{-\varepsilon/2}}
\end{equation}
where
\begin{equation}
\label{eqn:bd}
b_d\equiv {2-d\over 2^{3d/2}\pi^{d/2}}\Gamma(1-d/2)
=1/(4\pi)+{\cal O}(\varepsilon).
\end{equation}

It is convenient to define the dimensionless renormalised couplings
as
\begin{equation}
\label{dimless}
g_{Rjk}\equiv \lambda_{Rjk}\left({2\over D_j+D_k}\right)^{d/2}
(2\sigma)^{-\varepsilon/2}\qquad.
\end{equation}
As will be seen, these are the natural expansion parameters for the
renormalised perturbation expansion.

\subsection{Renormalisation of ${\cal O}^*$}
As discussed in the previous section, we are interested in computing the
asymptotic behaviour at large $t$ of
\begin{equation}
\label{eqn:overlineG}
\overline G_{\cal O}(t)
=\int\prod_{j=1}^p\prod_{\nu_j=1}^{n_j}d^d\!r_j^{\nu_j}
\langle\phi_j(t,{\bf r}_j^{\nu_j}){\cal O}^*(0,{\bf 0})\rangle\quad,
\end{equation}
in the regularised bare theory, where 
${\cal O}^*=\prod_{j=1}^p\big(\phi_j^*\big)^{n_j}$.
However, for the purposes of renormalising $\cal O^*$, it is more
convenient to choose the time arguments of the fields $\phi_j$ to be
independent, and to consider the laplace transform with respect to these
times. Thus we define
\begin{equation}
G_{\cal O}(\{t_j^{\nu_j}\})
=\int\prod_{j=1}^p\prod_{\nu_j=1}^{n_j}d^d\!r_j^{\nu_j}
\langle\phi_j(t_j^{\nu_j},{\bf r}_j^{\nu_j})
{\cal O}^*(0,{\bf 0})\rangle\quad,
\end{equation}
and
\begin{equation}
\tilde{G}_{\cal O}(\{s_j^{\nu_j}\})=
\int_{0}^{\infty} 
\prod_{j=1}^{p} \prod_{\nu_{j}=1}^{n_{j}} \left\{
dt_{j}^{\nu_{j}} \
e^{-s_{j}^{\nu_{j}} t_{j}^{\nu_{j}}} \right\}
G_{\cal O}(\{t_j^{\nu_j}\})
\end{equation}
{}From this we may define the irreducible vertex function\cite{Ami84},
by truncating the external propagators: 
\begin{equation}
\Gamma_{\cal O}(\{s_j^{\nu_j}\};\{\lambda_{jk}\})=
{\tilde G_{\cal O}(\{s_j^{\nu_j}\})\over
\prod_{j=1}^{p} \prod_{\nu_j=1}^{n_j}
\Big(s_j^{\nu_j}\Big)^{-1}}  .
\label{eqn:Gamma}
\end{equation}
The 
renormalised vertex function is
\begin{equation}
\Gamma_{{\cal O} \, R}(\{s_j^{\nu_j}\};
\{g_{Rjk}\}, \sigma) =
Z_{\cal O}(\{\lambda_{jk}\}, \sigma) 
\Gamma_{\cal O}(\{s_j^{\nu_j}\};\{\lambda_{jk}\})
\label{eqn:nGamma}
\end{equation}
where $Z_{\cal O}$ is fixed by the normalisation condition
\begin{equation}
\left. 
 \Gamma_{{\cal O} \, R}(\{s_j^{\nu_j}\};
\{g_{Rjk}\}, \sigma) 
\right|_{s_{1}^1=s_{1}^2=\cdots=s_{p}^{n_{p}}=\sigma} 
=1. 
\label{eqn:Z}
\end{equation}
In writing (\ref{eqn:nGamma}), we have made it clear that that the
(un)renormalised vertex function is to be thought of as depending on the
(un)renormalised couplings.

Now, although we have used the condition (\ref{eqn:Z})
on $\Gamma_{{\cal O}\,R}$ to define $Z_{\cal O}$, the 
\em same \em multiplicative renormalisation also renders 
\begin{equation}
\label{overlineGR}
\overline G_{{\cal O}\,R}(t;\{\lambda_{Rjk}\},\sigma)
=Z_{\cal O}\overline G_{{\cal O}}(t;\{\lambda_{jk}\})
\end{equation}
finite, for $t>0$, where $\overline G_{{\cal O}}$ is defined in
(\ref{eqn:overlineG}).
For this to be true, it is important
that the fields $\phi_j(t,{\bf r}_j^{\nu_j})$ are not evaluated at the
same point. This would lead to further UV divergences. However, these
occur on a set of measure zero in the integration in (\ref{eqn:overlineG}), 
and are harmless.

\subsection{Callan-Symanzik equation for $\overline G_{\cal O}$}

Define the RG functions
\begin{equation}
\beta_{jk}(g_{R \, jk}) = \sigma \left( \frac{\partial g_{R \, jk}}
{\partial \sigma} \right)_{\{\lambda_{jk}, D_{j}\}}.
\label{eqn:beta}
\end{equation}
and
\begin{equation}
\gamma_{\cal O}(\{g_{R\,jk}\}) = \left( \sigma
\frac{\partial}{\partial \sigma} 
\ln Z_{\cal O}(\{\lambda_{jk}\}, \sigma)
\right)_{\{\lambda_{jk}\}}.
\label{eqn:gamma}
\end{equation}
The fact that $\sigma(\partial/\partial\sigma)\overline G_{\cal
O}|_{\{\lambda_{jk}\}}=0$ then implies the Callan-Symanzik equation
\begin{equation}
\left(\sigma \frac{\partial}{\partial \sigma}
-\gamma_{{\cal O}}(\{g_{R \, jk}\})
+ \sum_{1 \leq j < k \leq p} 
\beta_{jk}(g_{R \, jk}) \frac{\partial}{\partial g_{R \, jk}} 
\right) 
\overline{G}_{{\cal O} R}(t, \{g_{R \, jk}\}, \sigma)=0.
\label{eqn:RGeq1}
\end{equation}
If the couplings $\{g_{R \, jk}\}$ flow towards a nontrivial fixed point
$\{g^*_{R \, jk}\}$
at which $\beta_{jk}(g^*_{R \, jk})=0$
(as we shall show  happens for $d<2$), then in estimating the leading
asymptotic behaviour as $\sigma\to\infty$ it is sufficient
to replace (\ref{eqn:RGeq1}) by
\begin{equation}
\left( \sigma \frac{\partial}{\partial \sigma}
-\gamma_{{\cal O}}^{*}
\right) 
\overline{G}_{{\cal O} R}(t, \sigma)=0,
\label{eqn:RGeq1f}
\end{equation}
where $\gamma_{\cal O}^*=\gamma_{\cal O}(\{g^*_{R\,jk}\})$.
This has the solution $\overline{G}_{{\cal O} R}(t, \sigma)\propto
\sigma^{\gamma_{\cal O}^*}$, as $\sigma\to\infty$ at fixed $t$.

However, simple dimensional analysis implies that 
$\overline{G}_{{\cal O} R}(t, \sigma)$ is function of only the
combination $(\sigma t)$. Hence we find that
\begin{equation}
\overline{G}_{{\cal O} R}(t, \sigma)\sim {\rm const.}\,t^{-\alpha}
\end{equation}
with
\begin{equation}
\label{eqn:alpha}
\alpha=-\gamma_{\cal O}^*\quad.
\end{equation}

\subsection{$\beta$-functions}
From (\ref{CCR}) and (\ref{dimless}), we find after some algebra 
(see App.~A)
\begin{equation}
\label{betafn}
\beta_{jk}(g_{R\,jk})=-\ffrac12(\varepsilon g_{R\,jk}
-b_dg_{R\,jk}^2).
\end{equation}
Note that this is exact to all orders in $g_{R\,jk}$, and that, for
$\varepsilon>0$, there is an infrared stable fixed point at
\begin{equation}
\label{eqn:g*}
g_{R\,jk}^*=\varepsilon/b_d=4\pi\varepsilon+{\cal O}(\varepsilon^2)\quad,
\end{equation}
whose value is independent of the diffusion constants.

\subsection{One-loop calculation of $\gamma_{\cal O}$}

\begin{figure}
\centerline{
\epsfxsize=4in
\epsfbox{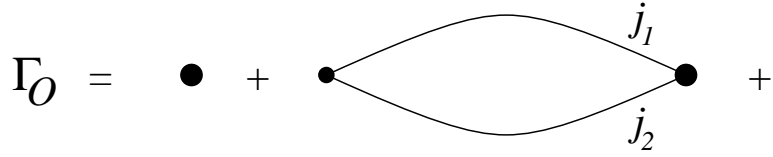}}
\caption{One-loop diagrams for $\Gamma_{\cal O}$.}
\label{fig1loop}
\end{figure}

Consider the expansion of the vertex function (\ref{eqn:Gamma}) 
as a power series in the coupling constants $\{\lambda_{jk}\}$. 
To first order, this is given by the sum of diagrams like that in
Fig.~\ref{fig1loop}, explicitly
\begin{equation}
\Gamma_{\cal O}
= 1 - \sum_{1 \leq j_{1} < j_{2} \leq p}
n_{j_{1}} n_{j_{2}} \lambda_{j_{1} j_{2}}
I_{1}(s_{j_{1}}^\mu, s_{j_{2}}^\nu;
D_{j_{1}}, D_{j_{2}})
+{\cal O} ( \{\lambda_{jk}^{2} \}),
\end{equation}
where the integral $I_{1}(s_{j_{1}}^\mu, s_{j_{2}}^\nu;
D_{j_{1}}, D_{j_{2}})$ is the same as occurs in the coupling
constant renormalisation: see Appendix~A. 
The combinatorial factor $n_{j_{1}} n_{j_{2}}$ counts the number of ways
the walkers from family $j_1$ can interact just once with those of
family $j_2$. The renormalisation constant $Z_{\cal O}$ is then the
inverse of this evaluated at the normalisation point
$s_j^{\nu_j}=\sigma$. Thus
\begin{equation}
\ln Z_{\cal O} =
\sum_{1 \leq j_{1} < j_{2} \leq p}
n_{j_{1}} n_{j_{2}} \lambda_{j_{1} j_{2}}
\frac{b_{d}}{\varepsilon} 
\left( \frac{2}{D_{j_{1}}+D_{j_{2}}} \right)^{d/2}
(2 \sigma)^{-\varepsilon/2} 
+{\cal O}(\{\lambda_{jk}^{2}\}),
\end{equation}
and so, by (\ref{eqn:gamma})
\begin{eqnarray}
\gamma_{\cal O}
&=& - \ffrac{1}{2} 
\sum_{1 \leq j_{1} < j_{2} \leq p}
n_{j_{1}} n_{j_{2}} \lambda_{j_{1} j_{2}} b_{d}
\left( \frac{2}{D_{j_{1}}+D_{j_{2}}} \right)^{d/2}
(2 \sigma)^{-\varepsilon/2}
+{\cal O}(\{\lambda_{jk}^{2}\}) \nonumber\\
&=&
-\ffrac{1}{2} 
\sum_{1 \leq j_{1} < j_{2} \leq p} 
n_{j_{1}} n_{j_{2}} \times
b_{d} g_{R \, j_{1} j_{2}} +{\cal O}(\{g_{R \,jk}^2\}).
\label{gammaO}
\end{eqnarray}
Next we set $g_{R \, j_1,j_2}=g^{*}$. By (\ref{eqn:g*}),
\begin{equation}
\gamma_{\cal O}^{*}= 
-\ffrac{1}{2} 
\sum_{1 \leq j_{1} < j_{2} \leq p} n_{j_{1}} n_{j_{2}} \
\varepsilon +{\cal O}(\varepsilon^2).
\end{equation}
Through (\ref{eqn:alpha}), this gives the result (\ref{eqn:result1})
up to ${\cal O}(\varepsilon)$.
We remark that to this order the result is independent of the diffusion
constants, as long as no pair $D_{j_1}+D_{j_2}$ vanishes.
This last is of course a pathological case, since then the two families
are immobile and cannot meet.

\subsection{Two-loop calculation}

\begin{figure}[h]
\centerline{
\epsfxsize=5in
\epsfbox{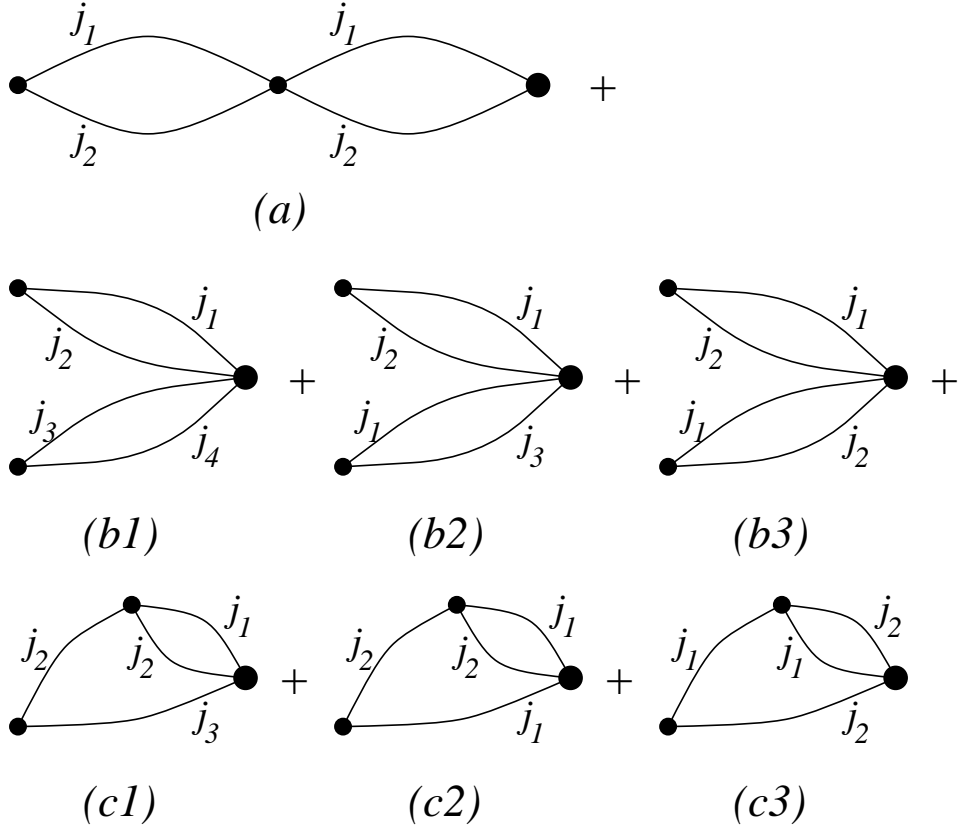}}
\caption{Two-loop contributions to $\Gamma_{\cal O}$.
Each diagram (with possible permutations of the labels) corresponds
to a term in (\ref{eqn:2loop}).}
\label{fig2loop}
\end{figure}

There are three types of diagrams contributing to $\Gamma_{\cal O}$ at
order $\lambda^2$. They are illustrated in Figs.~\ref{fig2loop}a-c. 
Contributions of types (a) and (b)
involve the one-loop integrals $I_1$ defined above,
while those of type (c) involve
\begin{eqnarray}
&& I_{2}(s_{j_1}^\mu, s_{j_2}^\nu, s_{j_3}^\rho;
D_{j_1}, D_{j_2}, D_{j_3})= \nonumber\\
&& \int\int {\big(d^d\!q/(2\pi)^d\big)\big(d^d\!k/(2\pi)^d\big)\over
\{ (s_{j_2}^\nu+s_{j_3}^\rho)
+(D_{j_2}+D_{j_3})({\bf k+q})^2 \}
\{(s_{j_1}^\mu+s_{j_2}^\nu+s_{j_3}^\rho)
+D_{j_1}{\bf k}^2+D_{j_2} {\bf q}^{2}+D_{j_3} ({\bf k+q})^2 \} }, \nonumber\\
\label{eqn:I2}
\end{eqnarray}
Define
\begin{eqnarray}
  \hat{I}_{1}(\sigma;D_{j_1}, D_{j_2}) &=&
  \left. I_{1}(s_{j_1}^\mu, s_{j_2}^\nu; D_{j_1}, D_{j_2})
  \right|_{s_{j_1}^\mu=s_{j_2}^\nu=\sigma},
\nonumber\\
  \hat I_{2}(\sigma; D_{j_1}, D_{j_2}, D_{j_3})
  &=& \left.I_{2}(s_{j_1}^\mu, s_{j_2}^\nu, s_{j_3}^\rho;
D_{j_1}, D_{j_2}, D_{j_3}) 
\right|_{s_{j_1}^\mu=s_{j_2}^\nu=s_{j_3}^\rho=\sigma}, 
\label{eqn:hatI}
\end{eqnarray}
Then
\begin{eqnarray}
\label{eqn:2loop}
Z_{\cal O}^{-1} = 
1&-& \sum_{1 \leq j_{1} < j_{2} \leq p}
n_{j_{1}} n_{j_{2}} \lambda_{j_{1} j_{2}}
\hat{I}_{1}(\sigma; D_{j_{1}}, D_{j_{2}} )  \nonumber\\
&+& \sum_{1 \leq j_{1} < j_{2} \leq p}
n_{j_{1}} n_{j_{2}} (\lambda_{j_{1} j_{2}} )^{2}
(\hat{I}_{1}(\sigma; D_{j_{1}}, D_{j_{2}}))^2 \nonumber\\
&+& \sum_{1 \leq j_{1} < j_{2} < j_{3} < j_{4} \leq p}
n_{j_{1}} n_{j_{2}} n_{j_{3}} n_{j_{4}}
\left\{ \lambda_{j_{1}j_{2}} \lambda_{j_{3}j_{4}}
\hat{I}_{1}(\sigma; D_{j_{1}}, D_{j_{2}})
\hat{I}_{1}(\sigma; D_{j_{3}}, D_{j_{4}}) \right. \nonumber\\
&& \hskip 1cm \left.
+\lambda_{j_{1}j_{3}} \lambda_{j_{2}j_{4}}
\hat{I}_{1}(\sigma; D_{j_{1}}, D_{j_{3}})
\hat{I}_{1}(\sigma; D_{j_{2}}, D_{j_{4}}) 
+\lambda_{j_{1}j_{4}} \lambda_{j_{2}j_{3}}
\hat{I}_{1}(\sigma; D_{j_{1}}, D_{j_{4}})
\hat{I}_{1}(\sigma; D_{j_{2}}, D_{j_{3}}) \right\} \nonumber\\
&+& \sum_{1 \leq j_{1} < j_{2} < j_{3} \leq p}
\left\{
n_{j_{1}}(n_{j_{1}}-1) n_{j_{2}} n_{j_{3}} 
\lambda_{j_{1} j_{2}} \lambda_{j_{1} j_{3}}
\hat{I}_{1}(\sigma; D_{j_{1}}, D_{j_{2}})
\hat{I}_{1}(\sigma; D_{j_{1}}, D_{j_{3}}) \right. \nonumber\\
&& \hskip 2cm 
+  n_{j_{1}} n_{j_{2}}(n_{j_{2}}-1) n_{j_{3}} 
\lambda_{j_{1} j_{2}} \lambda_{j_{2} j_{3}}
\hat{I}_{1}(\sigma; D_{j_{1}}, D_{j_{2}})
\hat{I}_{1}(\sigma; D_{j_{2}}, D_{j_{3}}) \nonumber\\
&& \hskip 2cm \left.
+ n_{j_{1}} n_{j_{2}} n_{j_{3}}(n_{j_{3}}-1)
\lambda_{j_{1} j_{3}} \lambda_{j_{2} j_{3}}
\hat{I}_{1}(\sigma; D_{j_{1}}, D_{j_{3}})
\hat{I}_{1}(\sigma; D_{j_{2}}, D_{j_{3}}) \right\} \nonumber\\
&+& \ffrac{1}{2} \sum_{1 \leq j_{1} < j_{2} \leq p}
n_{j_{1}}(n_{j_{1}}-1) n_{j_{2}} (n_{j_{2}}-1)
(\hat{I}_{1}(\sigma; D_{j_{1}}, D_{j_{2}}))^2 \nonumber\\
&+& \sum_{1 \leq j_{1} < j_{2} < j_{3} \leq p}
n_{j_{1}} n_{j_{2}} n_{j_{3}} \left\{
\lambda_{j_{1} j_{2}} \lambda_{j_{2} j_{3}}
\hat{I}_{2}(\sigma; D_{j_{1}}, D_{j_{2}}, D_{j_{3}}) 
+ \lambda_{j_{1} j_{2}} \lambda_{j_{1} j_{3}}
\hat{I}_{2}(\sigma; D_{j_{2}}, D_{j_{1}}, D_{j_{3}}) \right. 
\nonumber\\
&& \hskip 2cm 
+\lambda_{j_{1} j_{3}} \lambda_{j_{2} j_{3}}
\hat{I}_{2}(\sigma; D_{j_{1}}, D_{j_{3}}, D_{j_{2}}) 
+\lambda_{j_{1} j_{3}} \lambda_{j_{1} j_{2}} 
\hat{I}_{2}(\sigma; D_{j_{3}}, D_{j_{1}}, D_{j_{2}}) 
\nonumber\\
&& \hskip 2cm \left.
+\lambda_{j_{2} j_{3}} \lambda_{j_{1} j_{3}}
\hat{I}_{2}(\sigma; D_{j_{2}}, D_{j_{3}}, D_{j_{1}})
+\lambda_{j_{2} j_{3}} \lambda_{j_{1} j_{2}}
\hat{I}_{2}(\sigma; D_{j_{3}}, D_{j_{2}}, D_{j_{1}}) \right\}
\nonumber\\
&+& \sum_{1 \leq j_{1} < j_{2} \leq p}
n_{j_{1}}(n_{j_{1}}-1) n_{j_{2}}
(\lambda_{j_{1} j_{2}})^2 
\hat I_{2}(\sigma; D_{j_{1}}, D_{j_{2}}, D_{j_{1}}) \nonumber\\
&+&  \sum_{1 \leq j_{1} < j_{2} \leq p}
n_{j_{1}}n_{j_{2}} (n_{j_{2}}-1) 
(\lambda_{j_{1} j_{2}})^2 
\hat I_{2}(\sigma; D_{j_{2}}, D_{j_{1}}, D_{j_{2}})
+{\cal O}(\{\lambda_{jk}^{3}\})
\end{eqnarray}
Each term in this sum corresponds to a diagram of class $(a)$ to
$(c3)$ in Fig.~\ref{fig2loop}. 
The combinatorial factors, polynomials in the $n_j$,
count the number of ways different walkers from a given family can
contribute to each of these processes. (It is simplest to check these
factors for small values of $p$ and $n_j$.)
From Appendix~A
\begin{equation}
\hat{I}_{1}(\sigma; D_{j}, D_{k})
=\frac{b_{d}}{\varepsilon} 
\left( \frac{2}{D_{j}+D_{k}}\right)^{d/2}
(2 \sigma)^{-\varepsilon/2}.
\label{eqn:hatI1}
\end{equation}
Moreover, as shown in Appendix~B 
\begin{eqnarray}
&& \hat{I}_{2}(\sigma; D_{j}, D_{k}, D_{\ell}) \nonumber\\
&=& \left(\frac{2}{D_{j}+D_{k}}\right)^{d/2}
\left(\frac{2}{D_{k}+D_{\ell}}\right)^{d/2}
\left[ \frac{1}{2 \varepsilon^2}
-\frac{1}{4 \varepsilon} \ln R(D_{j}, D_{k}, D_{\ell})
+{\cal O}(1) \right] b_{d}^2 (2 \sigma)^{-\varepsilon}, 
\label{eqn:I2A}
\end{eqnarray}
where $b_{d}$ is given by (\ref{eqn:bd}) and
$R(D_{j}, D_{k}, D_{\ell})$ is given by (\ref{eqn:R0}). 

Next we compute $\ln Z_{\cal O}$ through 
${\cal O}(\{\lambda_{jk}^2\})$,
perform the differentiation $\sigma\partial/\partial\sigma$ at fixed
bare couplings $\lambda_{jk}$, then re-express the result as a series
in the $g_{R\,jk}$, to the same order. 
The result has the form
\begin{equation}
\gamma_{\cal O}(\{g_{R \, jk}\})
=B_{1}+\frac{1}{\varepsilon} B_{2} b_{d}^2+B_{3}
+{\cal O}(\{g_{R \, jk}^{3}\})
\end{equation}
with
\begin{eqnarray}
B_{1} &=& -\ffrac{1}{2} \sum_{1 \leq j_{1} < j_{2} \leq p}
n_{j_{1}} n_{j_{2}} b_{d} g_{R \, j_{1} j_{2}}, \nonumber\\
B_{2} &=& - \ffrac{1}{2} \sum_{1 \leq j_{1} < j_{2} \leq p}
n_{j_{1}} n_{j_{2}}
(g_{R \, j_{1} j_{2}})^2
+\sum_{1 \leq j_{1} < j_{2} \leq p}
n_{j_{1}} n_{j_{2}} 
(g_{R \, j_{1} j_{2}})^2
\nonumber\\
&& +\sum_{1 \leq j_{1} < j_{2} < j_{3} < j_{4} \leq p}
n_{j_{1}} n_{j_{2}} n_{j_{3}} n_{j_{4}} 
( g_{R \, j_{1} j_{2}} g_{R \, j_{3} j_{4}} 
+ g_{R \, j_{1} j_{3}} g_{R \, j_{2} j_{4}} 
+ g_{R \, j_{1} j_{4}} g_{R \, j_{2} j_{3}} ) \nonumber\\
&& +\sum_{1 \leq j_{1} < j_{2} < j_{3} \leq p} 
n_{j_{1}} n_{j_{2}} n_{j_{3}} \Big\{
(n_{j_{1}}-1) g_{R \, j_{1} j_{2}} g_{R \, j_{1} j_{3}} 
+ (n_{j_{2}}-1) g_{R \, j_{1} j_{2}} g_{R \, j_{2} j_{3}}  \nonumber\\
&& \hskip 4cm 
+ (n_{j_{3}}-1) g_{R \, j_{1} j_{3}} g_{R \, j_{2} j_{3}} \Big\}
\nonumber\\
&& + \ffrac{1}{2}
\sum_{1 \leq j_{1} < j_{2} \leq p}
n_{j_{1}}(n_{j_{1}}-1) n_{j_{2}}(n_{j_{2}}-1)
(g_{R \, j_{1} j_{2}})^2 \nonumber\\
&&+ \sum_{1 \leq j_{1} < j_{2} < j_{3} \leq p}
n_{j_{1}} n_{j_{2}} n_{j_{3}} 
( g_{R \, j_{1} j_{2}} g_{R \, j_{2} j_{3}}
+ g_{R \, j_{2} j_{3}} g_{R \, j_{1} j_{3}} 
+ g_{R \, j_{1} j_{3}} g_{R \, j_{1} j_{2}} )
\nonumber\\
\label{eqn:B2A}
&& +\ffrac{1}{2} \sum_{1 \leq j_{1} < j_{2} \leq p}
n_{j_{1}} n_{j_{2}} (n_{j_{1}}+n_{j_{2}}-2)
(g_{R \, j_{1} j_{2}})^2
- \ffrac{1}{2} 
\left( \sum_{1 \leq j_{1} < j_{2} \leq p}
n_{j_{1}} n_{j_{2}} g_{R \, j_{1} j_{2}} \right)^2, \\
B_{3} &=& -\ffrac{1}{2}
\sum_{1 \leq j_{1} < j_{2} < j_{3} \leq p}
n_{j_{1}} n_{j_{2}} n_{j_{3}} 
\Big\{ g_{R \, j_{1} j_{2}} g_{R \, j_{2} j_{3}} 
\ln R(D_{j_{1}}, D_{j_{2}}, D_{j_{3}}) \nonumber\\
&& \hskip 2cm
+g_{R \, j_{2} j_{3}} g_{R \, j_{1} j_{3}}
\ln R(D_{j_{2}}, D_{j_{3}}, D_{j_{1}}) 
+ g_{R \, j_{1} j_{3}} g_{R \, j_{1} j_{2}} 
\ln R(D_{j_{3}}, D_{j_{1}}, D_{j_{2}}) \Big\} b_{d}^2 
\nonumber\\
&& - \ffrac{1}{4}
\sum_{1 \leq j_{1} < j_{2} \leq p}
n_{j_{1}} n_{j_{2}} 
\Big\{(n_{j_{1}}-1) \ln R(D_{j_{1}}, D_{j_{2}}, D_{j_{1}}) 
+ (n_{j_{2}}-1) \ln R(D_{j_{2}}, D_{j_{1}}, D_{j_{2}}) \Big\}
b_{d}^2 (g_{R \, j_{1} j_{2}})^2. \nonumber
\end{eqnarray}

An important check of this calculation is that the double poles in 
$\varepsilon$ in the two-loop contributions are cancelled by the
coupling constant renormalisation in the one-loop contributions.
This has the consequence that $\gamma_{\cal O}(\{g_{R\,jk}\})$ has a
finite limit as $\varepsilon\to0$, that is, $B_2$ vanishes. This is
shown in Appendix C.

Now we set
\begin{equation}
g_{R \, jk}=g_{R}^{*}=\frac{\varepsilon}{b_{d}}
\qquad \mbox{for all} \quad
(j, k),
\end{equation}
following (\ref{eqn:g*}).
Then we have
\begin{equation}
\gamma_{\cal O}^{*}=B^{*}_{1} \varepsilon
+B^{*}_{3} \varepsilon^2 +{\cal O}(\varepsilon^2)
\end{equation}
with
\begin{eqnarray}
B_{1}^{*} &=& - \ffrac{1}{2} \sum_{1 \leq j_{1} < j_{2} \leq p}
n_{j_{1}} n_{j_{2}}  \nonumber\\
B_{3}^{*} &=& -\ffrac{1}{2} 
\sum_{1 \leq j_{1} < j_{2} < j_{3} \leq p}
n_{j_{1}} n_{j_{2}} n_{j_{3}} 
\Big\{ \ln R(D_{j_{1}}, D_{j_{2}}, D_{j_{3}})
+ \ln R(D_{j_{2}}, D_{j_{3}}, D_{j_{1}})
+ \ln R(D_{j_{3}}, D_{j_{1}}, D_{j_{2}}) \Big\}
\nonumber\\
&-& \ffrac{1}{4} 
\sum_{1 \leq j_{1} < j_{2} \leq p}
n_{j_{1}} n_{j_{2}} \Big\{(n_{j_{1}}-1)
\ln R(D_{j_{1}}, D_{j_{2}}, D_{j_{1}})
+(n_{j_{2}}-1)
\ln R(D_{j_{2}}, D_{j_{1}}, D_{j_{2}}) \Big\} \nonumber\\
&=& -\ffrac{1}{2} \sum_{1 \leq j_{1} < j_{2} < j_{3} \leq p}
n_{j_{1}} n_{j_{2}} n_{j_{3}}
\ln \left( \frac{(D_{j_{1}}D_{j_{2}}+D_{j_{1}}D_{j_{3}}
+D_{j_{2}}D_{j_{3}})^3}
{(D_{j_{1}}+D_{j_{2}})^2 (D_{j_{2}}+D_{j_{3}})^2
(D_{j_{3}}+D_{j_{1}})^2} \right) \nonumber\\
&-& \ffrac{1}{4} 
\sum_{1 \leq j_{1} < j_{2} \leq p}
n_{j_{1}} n_{j_{2}} \Big\{ n_{j_{1}}
\ln R(D_{j_{1}}, D_{j_{2}}, D_{j_{1}})
+ n_{j_{2}}
\ln R(D_{j_{2}}, D_{j_{1}}, D_{j_{2}}) \Big\} \nonumber\\
&+& \ffrac{1}{4} 
\sum_{1 \leq j_{1} \leq j_{2} \leq p}
n_{j_{1}} n_{j_{2}} \ln \left(
\frac{D_{j_{1}}D_{j_{2}}(D_{j_{1}}+2D_{j_{2}})
(2 D_{j_{1}}+D_{j_{2}})}
{(D_{j_{1}}+D_{j_{2}})^4} \right). \nonumber
\end{eqnarray}
Through (\ref{eqn:alpha}), this gives the result (\ref{eqn:result1}).
\subsection{Two Dimensions}
When  $d=2$ the couplings $\{g_{Rjk}\}$ are marginally irrelevant, that
is they flow logarithmically slowly towards zero. In that case, it is
not sufficient to set them equal to their fixed-point values, but
instead the full Callan-Symanzik equation (\ref{eqn:RGeq1}) must be solved.
Using the fact that $\overline G_{{\cal O}R}(t,\{g_{Rjk}\},\sigma)$
depends on $\sigma$ and $t$ only through combination $(\sigma t)$,
this may be rewritten
\begin{equation}
\left(t\frac{\partial}{\partial t}
-\gamma_{{\cal O}}(\{g_{R \, jk}\})
+ \sum_{1 \leq j < k \leq p}
\beta_{jk}(g_{R \, jk}) \frac{\partial}{\partial g_{R \, jk}}
\right)
\overline{G}_{{\cal O} R}(t, \{g_{R \, jk}\}, \sigma)=0.
\end{equation}
The solution by the method of characteristics is standard.
Define
running couplings $\{\tilde g_{jk}(u)\}$ by
\begin{equation}
u{d\over du}\tilde g_{jk}(u)=-\beta_{jk}(\tilde g_{jk}(u))
\end{equation}
with initial conditions $\tilde g_{jk}(1)=g_{R\,jk}$. 
Then
\begin{equation}
\label{eqn:CSsoln}
\overline G_{{\cal O} R}(t, \{g_{R\,jk}\}, \sigma)
={\rm e}^{\int_1^{\sigma t}\gamma_{\cal O}(\{\tilde g_{R\,jk}\})(du/u)}\,
\overline G_{{\cal O} R}(\sigma^{-1}, \{\tilde g_{jk}(\sigma t)\}, \sigma)
\end{equation}
In our case, 
\begin{equation}
\tilde g_{jk}(u)={1\over \ffrac12b_2\ln u+g_{R\,jk}^{-1}}=
{2\over b_2\ln u}+{\cal O}((\ln u)^{-2})
\end{equation}
so that, using (\ref{gammaO}) the exponent in (\ref{eqn:CSsoln}) is
\begin{equation}
\ln\ln(\sigma t)\sum_{1\leq j_1<j_2\leq p}n_{j_1}n_{j_2}
+{\cal O}((\ln(\sigma t))^{-1})
\end{equation}
Exponentiating this yields the
result quoted in (\ref{2dlog}). The dependence of
the last factor in (\ref{eqn:CSsoln}) on $\{\tilde g_{jk}(\sigma t)\}$
also generates corrections which are down by ${\cal O}((\ln(\sigma
t))^{-1}$. All the non-universal behaviour resides in these, and higher
order, corrections. Note the absence of corrections
${\cal O}(\ln\ln t/\ln t)$, which may be traced to the lack of
higher-order terms in the beta-functions (\ref{betafn}).
\setcounter{equation}{0}
\section{Discussion}
\label{sec5}
We have presented a generalisation of the vicious walker problem 
in which walkers from different families annihilate on meeting, but
walkers from the same family ignore each other. We have studied the
problem in a field-theoretic renormalisation group framework, suitable
for understanding universal quantities such as critical exponents. We
have focussed on the probability that all walkers have survived up
to time $t$, and we have showed that, in dimension $d<2$, this decays
as a power, $t^{-\alpha(\{n_j\})}$, where $n_j$ is the number of walkers
in the $j$th family. While this result is true to all orders in
$\varepsilon\equiv2-d$, the actual values of the exponents can, by this
method, be evaluated only as a power series in $\varepsilon$, which we
have carried out to second order. The coefficient of the ${\cal
O}(\varepsilon^2)$ term depends on the ratios of the diffusivities of
each family, as well as the $\{n_j\}$. 
The lack of dependence of the ${\cal O}(\varepsilon)$ term on this ratio
may be traced mathematically to the fact that the same bubble diagram
enters into the coupling constant renormalisation (Fig.~\ref{figcc}) and
the renormalisation of the composite operator ${\cal O}^*$
(Fig.~\ref{fig1loop}). 

For the same reason, the exponent for $N=2$ walkers does not depend
on the ratio of their diffusivities, because in this case the one-loop
result is correct to all orders (as long as we do not expand the 
coefficients in powers of $\varepsilon$.)
The same would be true, to first order for $N>k$, and to all orders
for $N=k$, if the two-body
interactions we consider were generalised to $k$-body interactions
with $k>2$,
although in this case the upper critical dimension would be reduced
to $d_c(k)=2/(k-1)$.

The value of $\alpha(\{n_j\})$ was shown to be related to the anomalous
dimension of a certain composite operator ${\cal O}^*_{\{n_j\}}
=\prod_j{\phi^*_j}^{n_j}$. This
structure means that our results can be straightforwardly extended
to other physical observables. For example, the \em reunion \em
probability $R(t)$ that all the walkers have survived up to time $t$ and
are all located within a distance ${\cal O}(b)\ll t^{1/2}$ 
of each other (where $b$
is for example the lattice spacing), is 
related to a correlation function
\begin{equation}
R(t)\sim\int\langle{\cal O}(t,{\bf r}){\cal O}^*(0,{\bf
0})\rangle d^d\!r
\end{equation}
where ${\cal O}\equiv\prod_j\phi_j^{n_j}$.
Since the theory is symmetric under
$(t\to-t,\phi_j\leftrightarrow\phi^*_j)$, 
$\cal O$ has the same anomalous dimension as ${\cal O}^*$.
Hence
\begin{equation}
R(t)\sim {\rm const.\ }t^{-(N-1)(d/2)-2\alpha(\{n_j\})}
\end{equation}
where the first term in the exponent comes from simple power counting,
and the factor of 2 in the 
second reflects the important fact that the anomalous scaling of
composite operators at different times (and points) is multiplicative. 

The fact that ${\cal O}^*_{\{n_j\}}$ is symmetric under
permutations of the families $j=1,\ldots,p$ has the important
consequence that the exponents $\alpha(\{n_j\})$ are also symmetric.
The form of the $\varepsilon$-expansion implies
that this is true to all orders in $\varepsilon$. It may be traced to
the operator product expansion (\ref{OPE}): the next-to-leading terms on
the right-hand side, which do not vanish on integration over the spatial
coordinates must contain at least two derivatives, for example
\begin{equation}
({\bf\nabla}\phi_1\cdot{\bf\nabla}\phi_2)\phi_1^{n_1-1}
\phi_2^{n_2-1}\prod_{3\leq j\leq p}\phi_j^{n_j}
\end{equation}

Power-counting shows that, at $d=2$, the contribution of such terms
is at least ${\cal O}(t^{-1})$ down on the leading term, and therefore, for
sufficiently small $\varepsilon$, they yield only corrections to the
leading behaviour which we have computed. However, since each term in
(\ref{OPE}) is renormalised separately, each gives rise to an independent
scaling exponent.
It is a very interesting question whether the first term in the OPE
is dominant all the way down to $d=1$. 
Such a result would imply that the asymptotic exponents (but not
necessarily the prefactors) for the
cases $\{n_1,n_2,n_3\}=\{2,1,1\}$ and $\{1,2,1\}$, for example, are
equal. Yet, in one
dimension, the two problems are certainly not isomorphic, because the
ordering of the families along the real line makes a difference.
A very similar
situation occurs in the problem of intersections of families of
Brownian paths in two dimensions: in that case the exponents have been
computed exactly\cite{Dup,LSW} and are known to be symmetric, despite the
fact that the ordering of the families around the annulus is a priori
relevant. In this example, this symmetry is also suggested by the
$\varepsilon$-expansion below the upper critical dimension, in this case
$d_c=4$ \cite{BPepsilon}.
However, a recent result of Bray and Blythe\cite{BB02} suggests that the
situation in $d=1$ is not so straightforward for our problem. 
In their Eq.~(4) they report a result for the survival exponent
for a single lamb, with diffusion constant $D'$, with $N_L$ lions to the
left and $N_R$ lions to the right. The diffusion constant of the lions is
$D$,
and their result is an expansion to first order in $D'/D$.
It is reported to depend on the
asymmetry $N_L-N_R$ as well as the sum $N_L+N_R$, while our analysis 
shows that it should depend only on the sum, to all orders in the
$\varepsilon$-expansion.

There are at least two possible resolutions:
\begin{enumerate}
\item the qualitative conclusions of the $\varepsilon$-expansion break
down in $d=1$, either because the non-leading terms in the OPE dominate,
or through some more systemic failure;
\item for $d=1$ there is a qualitative difference between the case
when walkers if different families strictly cannot
pass each other, and that, more appropriate to the field theory
approach, when the annihilation rate is finite and therefore the order
along the real line is not preserved.
\end{enumerate}

In any case, 
it may be shown, by generalising the arguments of Krapivsky and 
Redner\cite{KR96,RK99,R01}, 
that for infinite annihilation rate in
$d=1$ the exponents $\alpha(\{n_j\})$ (and
indeed all the non-leading exponents), are simply related to the
eigenvalues of the Dirichlet problem in a certain compact region consisting
of a spherical hyperpolygon on the sphere $S_{N-2}$.
Thus, if the postulated symmetry were to hold, it would imply
that the Laplacians in different regions which are related by
permutations of the $\{n_j\}$ are isospectral. We plan to address this
question in a future publication. 

\vspace{5mm}
\noindent\em Acknowledgements\em.
JC was supported in part by the Engineering and
Physical Sciences Research Council under Grant GR/J78327.
This work was done during a stay of MK in 
Theoretical Physics, Oxford University.
MK would like to thank the department for hospitality.

\appendix
\setcounter{equation}{0}
\section{Coupling constant renormalisation}

\begin{figure}
\centerline{
\epsfxsize=5in
\epsfbox{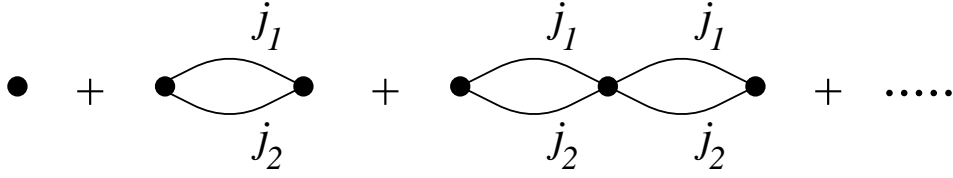}}
\caption{Diagrams renormalising $\lambda_{j_1j_2}$.}
\label{figcc}
\end{figure}

The diagrams contributing to $\Gamma_{j_1j_2}^{(2,2)}$ are shown in
Fig.~\ref{figcc}. 
In $(s,{\bf q})$ space they give a geometric sum.
We have therefore the exact result
\begin{equation}
\Gamma_{j_1j_2}^{(2,2)}\big((s_{j_1}',{\bf q}_{j_1}'),(s_{j_2}',{\bf q}_{j_2}')
;(s_{j_1},{\bf q}_{j_1}),(s_{j_2},{\bf q}_{j_2})\big)=
{\lambda_{j_1j_2}\over
1+\lambda_{j_1j_2}\int{d^d\!k\over(2\pi)^d}
{1\over s+D_{j_1}{\bf k}^2+D_{j_2}({\bf q}-{\bf k})^2}}
\end{equation}
where $s\equiv s_{j_1}'+s_{j_2}'=s_{j_1}+s_{j_2}$ 
and ${\bf q}\equiv{\bf q}_{j_1}'+{\bf q}_{j_2}'=
{\bf q}_{j_1}+{\bf q}_{j_2}$. The renormalised coupling is the value of this 
at $s_{j_1}'=s_{j_2}'=s_{j_1}=s_{j_2}=\sigma$ and 
${\bf q}_{j_1}'={\bf q}_{j_2}'={\bf q}_{j_1}={\bf q}_{j_2}=0$.
Thus
\begin{equation}
\lambda_{R\,j_1j_2}=
{\lambda_{j_1j_2}\over 
1+\lambda_{j_1j_2}\hat I_1(\sigma;D_{j_1},D_{j_2})}
\end{equation}
where 
\begin{eqnarray}
\hat I_1(\sigma;D_{j_1},D_{j_2})&=& 
\int{d^d\!k\over(2\pi)^d}\int_0^\infty d\alpha 
{\rm e}^{-\alpha(2\sigma+(D_{j_1}+D_{j_2}){\bf k}^2)}\\
&=&{1\over(2\pi)^d}\left({\pi\over D_{j_1}+D_{j_2}}\right)^{d/2}
\int_0^\infty d\alpha \alpha^{-d/2}{\rm e}^{-\alpha(2\sigma)}\\
&=&{\Gamma(1-d/2)\over2^{3d/2}\pi^{d/2}}
\left({2\over D_{j_1}+D_{j_2}}\right)^{d/2}(2\sigma)^{-\varepsilon/2}
\end{eqnarray}

The dimensionless coupling (\ref{dimless}) is then given as
\begin{equation}
g_{R \, j_1j_2}
=\frac{\displaystyle{ \lambda_{j_1j_2} 
\left(\frac{2}{D_{j_1}+D_{j_2}} \right)^{d/2}
(2\sigma)^{-\varepsilon/2}}}
{\displaystyle{1+\frac{b_{d}}{\varepsilon}
\lambda_{j_1j_2} \left(\frac{2}{D_{j_1}+D_{j_2}}\right)^{d/2}
 (2\sigma)^{-\varepsilon/2}}}.
\label{eqn:gR1}
\end{equation}
Thus
\begin{equation}
\lambda_{j_1j_2}
\left(\frac{D_{j_1}+D_{j_2}}{2} \right)^{-d/2}(2 \sigma)^{-\varepsilon/2}
= \frac{g_{R \, j_1j_2}}
{\displaystyle{1-g_{R \, j_1j_2} \frac{b_{d}}{\varepsilon}}}
\label{eqn:gR2}
\end{equation}
Differentiating this equation with respect to $\sigma$ at
fixed $(\lambda_{j_1j_2},D_{j_1},D_{j_2})$ and using the definition of the
beta-function (\ref{eqn:beta}) 
\begin{equation}
(-\varepsilon/2){g_{Rj_1j_2}\over1-g_{Rj_1j_2}{b_d\over\varepsilon}}
=\left({1\over 1-g_{Rj_1j_2}{b_d\over\varepsilon}}\right)^2
\beta_{j_1j_2}(g_{Rj_1j_2})
\end{equation}

That is,
\begin{equation}
\beta_{j_1j_2}(g_{Rj_1j_2})= -\ffrac12(\varepsilon g_{Rj_1j_2}- 
b_dg_{Rj_1j_2}^2).
\label{eqn:beta2}
\end{equation}

\setcounter{equation}{0}
\section{The integral $\hat I_2$}
At the normalisation point, the two-loop diagram Fig.~\ref{fig2loop}$c$ leads
to the integral
\begin{eqnarray}
&&\hat I_2(\sigma;D_1,D_2,D_3)\nonumber\\
&=&\int{d^d\!q\over(2\pi)^d}\int{d^d\!k\over(2\pi)^d} 
{1\over\big(2\sigma+(D_2+D_3){\bf k}^2\big)
\big(3\sigma+D_1{\bf q}^2+D_2({\bf k}+{\bf q})^2+D_3{\bf k}^2\big)}
\end{eqnarray}
As usual, the denominators may be combined using a Feynman parameter
integration over $x$:
\begin{equation}
\hat I_2(\sigma;D_1,D_2,D_3)=
\int_0^1dx\int{d^d\!q\over(2\pi)^d}\int{d^d\!k\over(2\pi)^d}
{1\over\big((D_2+D_3){\bf k}^2+x(D_1+D_2){\bf q}^2+
2xD_2{\bf k}\cdot{\bf q}\big)^2}
\end{equation}
The wave-number integrals are now standard, and yield
\begin{equation}
\hat I_2={\pi^d\over(2\pi)^d}\Gamma(2-d)(D_1+D_2)^{-d/2}(D_2+D_3)^{-d/2}
\sigma^{d-2}\,\times\,J
\end{equation}
where
\begin{equation}
J=\int_0^1dx\,x^{-d/2}(2+x)^{d-2}\left(1-{xD_2^2\over
(D_1+D_2)(D_2+D_3)}\right)^{-d/2}
\end{equation}
This has a simple pole at $d=2$, arising from the end-point at $x=0$.
However, we also have to extract the finite part. This may be done by 
writing $J=J_1+J_2$ where
\begin{eqnarray}
J_1&=&\int_0^1dx\,x^{-d/2}\,2^{d-2}={2\over\varepsilon}\,2^{d-2}\\
J_2&=&\int_0^1dx\,x^{-d/2}
\left[(2+x)^{d-2}\left(1-{xD_2^2\over
(D_1+D_2)(D_2+D_3)}\right)^{-d/2}-2^{d-2}\right]
\end{eqnarray}
The second integral is finite at $d=2$:
\begin{eqnarray}
J_2&=&\int_0^1{dx\over x}\left[\left(1-{xD_2^2\over
(D_1+D_2)(D_2+D_3)}\right)^{-1}-1\right]+{\cal O}(\varepsilon)
\nonumber\\
&=&-\ln R(D_1,D_2,D_3)+{\cal O}(\varepsilon)
\end{eqnarray}
where
\begin{equation}
R(D_1,D_2,D_3)\equiv {D_1D_2+D_2D_3+D_1D_3\over(D_1+D_2)(D_2+D_3)}
\end{equation}
Recalling the definition (\ref{eqn:bd}) of $b_d$, we therefore find, after
some algebra, 
\begin{equation}
{\hat I_2\over b_d^2}=\left({2\over D_1+D_2}\right)^{d/2}
\left({2\over D_2+D_3}\right)^{d/2}
(2\sigma)^{-\varepsilon}\left[{1\over2\varepsilon^2}-
{1\over4\varepsilon}\ln R+{\cal O}(1)\right]
\end{equation}
which leads directly to (\ref{eqn:I2A}).
It should be remarked that a crucial simplification in this calculation
arises because
\begin{equation}
{\Gamma(2-d)\over\Gamma(1-d/2)^2}=\ffrac\varepsilon 4\big(
1+{\cal O}(\varepsilon^2)\big)\quad.
\end{equation}
\setcounter{equation}{0}
\section{Verification that $B_2=0$}

By (\ref{eqn:B2A}), 
\begin{eqnarray}
&& B_{2} =
\ffrac{1}{2} \sum_{1 \leq j_{1} < j_{2} \leq p}
n_{j_{1}} n_{j_{2}} 
(g_{R \, j_{1} j_{2}})^2
\nonumber\\
&& +\sum_{1 \leq j_{1} < j_{2} < j_{3} < j_{4} \leq p}
n_{j_{1}} n_{j_{2}} n_{j_{3}} n_{j_{4}} 
( g_{R \, j_{1} j_{2}} g_{R \, j_{3} j_{4}} 
+ g_{R \, j_{1} j_{3}} g_{R \, j_{2} j_{4}} 
+ g_{R \, j_{1} j_{4}} g_{R \, j_{2} j_{3}} ) \nonumber\\
&& +\sum_{1 \leq j_{1} < j_{2} < j_{3} \leq p} 
n_{j_{1}} n_{j_{2}} n_{j_{3}}
(n_{j_{1}}g_{R \, j_{1} j_{2}} g_{R \, j_{1} j_{3}} 
+ n_{j_{2}} g_{R \, j_{1} j_{2}} g_{R \, j_{2} j_{3}} 
+ n_{j_{3}} g_{R \, j_{1} j_{3}} g_{R \, j_{2} j_{3}})
\nonumber\\
&& -\sum_{1 \leq j_{1} < j_{2} < j_{3} \leq p}
n_{j_{1}} n_{j_{2}} n_{j_{3}} 
(g_{R \, j_{1} j_{2}} g_{R \, j_{1} j_{3}} 
+ g_{R \, j_{1} j_{2}} g_{R \, j_{2} j_{3}} 
+ g_{R \, j_{1} j_{3}} g_{R \, j_{2} j_{3}})
\nonumber\\
&& + \ffrac{1}{2}
\sum_{1 \leq j_{1} < j_{2} \leq p}
n_{j_{1}}^2 n_{j_{2}}^2
(g_{R \, j_{1} j_{2}})^2 
- \ffrac{1}{2}
\sum_{1 \leq j_{1} < j_{2} \leq p}
n_{j_{1}} n_{j_{2}} (n_{j_{1}}+n_{j_{2}})
(g_{R \, j_{1} j_{2}})^2 
 + \ffrac{1}{2}
\sum_{1 \leq j_{1} < j_{2} \leq p}
n_{j_{1}} n_{j_{2}}
(g_{R \, j_{1} j_{2}})^2 \nonumber\\
&&+ \sum_{1 \leq j_{1} < j_{2} < j_{3} \leq p}
n_{j_{1}} n_{j_{2}} n_{j_{3}} 
( g_{R \, j_{1} j_{2}} g_{R \, j_{2} j_{3}}
+g_{R \, j_{2} j_{3}} g_{R \, j_{1} j_{3}}
+ g_{R \, j_{1} j_{3}} g_{R \, j_{1} j_{2}})
\nonumber\\
&& +\ffrac{1}{2} \sum_{1 \leq j_{1} < j_{2} \leq p}
n_{j_{1}} n_{j_{2}} ( n_{j_{1}}+n_{j_{2}}) 
(g_{R \, j_{1} j_{2}})^2
- \sum_{1 \leq j_{1} < j_{2} \leq p}
n_{j_{1}}n_{j_{2}} (g_{R \, j_{1} j_{2}})^2
 - \ffrac{1}{2} 
\left( \sum_{1 \leq j_{1} < j_{2} \leq p}
n_{j_{1}} n_{j_{2}} g_{R \, j_{1} j_{2}} \right)^2
\nonumber\\
&=& 
\sum_{1 \leq j_{1} < j_{2} < j_{3} < j_{4} \leq p}
n_{j_{1}} n_{j_{2}} n_{j_{3}} n_{j_{4}} 
( g_{R \, j_{1} j_{2}} g_{R \, j_{3} j_{4}} 
+ g_{R \, j_{1} j_{3}} g_{R \, j_{2} j_{4}} 
+ g_{R \, j_{1} j_{4}} g_{R \, j_{2} j_{3}} ) \nonumber\\
&& +\sum_{1 \leq j_{1} < j_{2} < j_{3} \leq p} 
n_{j_{1}} n_{j_{2}} n_{j_{3}}
(n_{j_{1}} g_{R \, j_{1} j_{2}} g_{R \, j_{1} j_{3}} 
+ n_{j_{2}} g_{R \, j_{1} j_{2}} g_{R \, j_{2} j_{3}}
+ n_{j_{3}} g_{R \, j_{1} j_{3}} g_{R \, j_{2} j_{3}} )
\nonumber\\
&& + \ffrac{1}{2}
\sum_{1 \leq j_{1} < j_{2} \leq p}
n_{j_{1}}^2 n_{j_{2}}^2
(g_{R \, j_{1} j_{2}})^2 
 - \ffrac{1}{2} 
\left( \sum_{1 \leq j_{1} < j_{2} \leq p}
n_{j_{1}} n_{j_{2}} g_{R \, j_{1} j_{2}} \right)^2
\nonumber\\
&=& 0.
\end{eqnarray}
\hfill \qed

\end{document}